# Priyadarshini Engineering College

(Approved by AICTE, New Delhi and Permanently Affiliated to Anna University, Chennai)
Chettiyappanur Village & Post, Vaniyambadi-635751, Vellore District, Tamil Nadu, India.
Listed in 2(f) & 12(B) Sections of UGC.

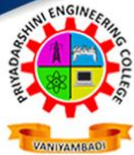
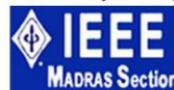

Technical Sponsor by
IEEE MADRAS Section

## CERTIFICATE

### International Conference on Electrical, Electronics, Computers, Communication, Mechanical and Computing (EECCMC) – 2018

This is to certify that **Mr. Seyyedmostafa Mousavi Janbehsarayi** has presented a paper entitled: **Dynamic Analysis of Tapping-mode AFM with Sidewall Probe Subjected to Effects of Probe Mass and Sidewall Extension** with paper code: **01-2018-401** in International Conference on Electrical, Electronics, Computers, Communication, Mechanical and Computing (EECCMC) – 2018, with catalog "CFP18O37-PRJ: 978-1-5386-4303-7", organized by Priyadarshini Engineering College, Vellore District, Tamil Nadu, India during 28th & 29th January 2018.

Certificate Proof

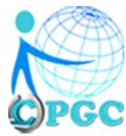

Dr. Siva Ganesh Malla
Director, CPGC

Dr. P. Natarajan
Principal, PEC

# CERTIFICATE

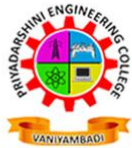

# Priyadarshini Engineering College

(Approved by AICTE, New Delhi and Permanently Affiliated to Anna University, Chennai)
Chettiyappanur Village & Post, Vaniyambadi-635751, Vellore District, Tamil Nadu, India.
Listed in 2(f) & 12(B) Sections of UGC.

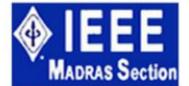

Technical Sponsor by
IEEE MADRAS Section

## International Conference on Electrical, Electronics, Computers, Communication, Mechanical and Computing (EECCMC) - 2018

This is to certify that **Mr. Seyyedmostafa Mousavi Janbehsarayi** has published a paper entitled: **Dynamic Analysis of Tapping-mode AFM with Sidewall Probe Subjected to Effects of Probe Mass and Sidewall Extension** with paper code: **01-2018-401** in International Conference on Electrical, Electronics, Computers, Communication, Mechanical and Computing (EECCMC) - 2018, with catalog "CFP18O37-PRT: 978-1-5386-4303-7", organized by Priyadarshini Engineering College, Vellore District, Tamil Nadu, India during 28th & 29th January 2018.

IEEE Conference Record # 43456

Certificate Proof

Dr. Siva Ganesh Malla
Director, CPGC

Dr. P. Natarajan
Principal, PEC

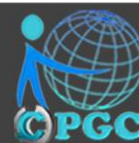

©PGC

# Dynamic analysis of tapping-mode AFM with sidewall probe subjected to effects of probe mass and sidewall extension

Sina Eftekhar

*Department of Mechanical Engineering, College of Engineering, University of Tehran, Tehran, Iran*

SeyyedMostafa Mousavi Janbeh Sarayi

*Department of Mechanical Engineering, College of Engineering, University of Tehran, Tehran, Iran*

*Abstract*— Atomic Force Microscopy with SideWall (AFM-SW) is widely used for nano-scale surface measurements at side surfaces. In the current study, by taking into consideration the effects of sidewall beam and its probe, an analytical method is developed to explore the dynamics of AFM-SW. The effect of probe mass, sidewall extension length, and tip sample interactions on the resonance frequencies and amplitude of Micro-Cantilever is widely investigated. The obtained results of the analytical model demonstrate the significant effect of these parameters on the dynamics of AFM-SW. To verify the accuracy of the analytical model, the obtained results are compared against the simulation data of previously published works and a good agreement is observed. Resonance Frequency (RF) of cantilever clearly declines when the mass of probe is taken to account, especially in higher RFs. Besides, probe effect on RF is higher when sidewall beam is longer. Resonance frequency decreases when tip-sample interaction or probe mass is high, yet the amount of reduction is intensified when probe mass and interaction together are at higher point.

An analytical method is developed to explore the dynamics of Atomic Force Microscopy with considering SideWall beam effects (AFM-SW). The effect of probe mass, sidewall extension length, and tip sample interactions on vibration of micro-cantilever is investigated. The obtained results are compared with previous literatures. The results show that Resonance Frequency (RF) of cantilever declines when the mass of probe is taken to account. Besides, probe effect on RF is higher when sidewall beam is longer. Resonance frequency decreases when tip-sample interaction or probe mass is high, yet the amount of reduction is intensified when they are at higher point.

*Keywords—Atomic Force microscopy; Sidewall Probe; Micro-Cantilever; Vibration; Resonant frequency*

## I. INTRODUCTION

Atomic force microscopy (AFM) is widely used in materials science and has found many applications in surface measurements. AFM is the latest scientific achievement in the field of imaging, topography, and manipulating structures at the nanometer scale in the past twenty years [1, 2, 3]. The AFM can be utilized to image the topography of materials in nano scales. Furthermore, the AFM is able to probe mechanical properties including obtaining an elastic modulus of a surface, and measuring modulus variations across a sample surface. Whereas scanning tunneling microscope was unusable for nonconductive samples, AFM was developed in 1986 to do away with the need for a conductive sample.

AFM is composed of electro-mechanical components using electro-mechanical techniques on an atomic scale [1, 2]. A clamped micro-beam with pyramidal or conical tip has an imperative role in the performance of AFM. The tip is usually made of silicon and nano-tubes with the tip head radius in the range of 0.7-5.0 nm. In reality, the dynamics behavior of AFM Micro-Cantilever system and its probe are sophisticated and have impressive effects on the AFM operations. Therefore, researchers and scientists have then been attracted to the AFM dynamics studies.

AFM with Common probe (C-AFM) play the main role in nano-scale surface measurement, but in scanning the sidewall and edge surfaces using the Common AFMs are not appropriate [4, 5, 6]. Scanning the sidewall and edge surfaces are denoted in micro/nano electromechanical systems (NEMS/MEMS) such as determining the waviness or roughness of the micro-structures e.g. micro-gears, micro injection nozzles, and integrated circuit (CI) structures [5, 7, 8]. Hence, the modified AFM with sidewall scanning capacity seemed to be necessary.

Dai et al. [4] developed an AFM probe in order to be applicable for sidewall scanning. The measurements were taken at the sidewalls microstructures e.g. microgears, microtrenches, and line edge roughness samples. The nano-scale surface measurements at sidewalls are really indispensable.

A new method of scanning the sidewalls in micro-nano structures has been presented [5]. The varieties of Assembled Cantilever Probe (ACP), related to the sidewall surface measurements, were studied and performed.

Chang et al. [7] studied sensitivity of the flexural (vertical) mode and the Resonant Frequencies (RFs) of AFM with SideWall probe (AFM-SW). It was shown that vibration sensitivity of AFM-SW is more in case of weak contact stiffness of tip-sample interaction, especially in the first RFs. However, inversely in the case of strong contact stiffness the higher RFs are more sensitive. They found that by increasing the interaction contact stiffness, the RFs increase. The effects of vertical extension length on the sensitivity and RFs of AFM-SW were also shown.

The sensitivity and RFs of ACP AFM was also investigated by Kahrobiyan et al. on both flexural and torsional RFs, using the Reyleigh-Ritz method [8]. The effects of contact stiffness and geometrical properties of assembled cantilever on the vertical/torsional RFs and its sensitivities were studied. The first vibration modes are more sensitive for low values of the contact stiffness, but the situation is reversed for high values. H. N.

Pishkenari et al. [9] used finite-element method (FEM) to investigate the influence of tip mass on tip-sample interaction forces. Eslami and Jalili [10] Presented a comprehensive analytical model for the AFM system using a distributed-parameters model of micro-cantilever beams utilized in AFM systems.

In most of the recent studies, obtaining the motion behavior of Micro-Cantilever (MC) e.g. Resonance Frequency (RF), amplitude, and vibration sensitivity have been investigated e.g. [3, 7, 8, 11, 12, 13, 14, 15, 16, 17, 18] whereas a closer and more accurate examination on the dynamic behavior of AFM micro-cantilever seems essential. In this study, the analytical method is performed by considering these effective parameters: (1) the tip mass (2) damping coefficient of MC and (3) visco-elastic forces between probe and sample in effective directions of normal and tangential. Considering and analyzing all the mentioned effective parameters, which have been neglected in previous researches, leads to obtaining more accurate response functions of MC oscillation. In this paper, resonance frequency and amplitude of vertical vibration of an AFM-SW have been studied taking into account the effects of probe mass, side wall length, and tip-sample effects.

## II. ANALYTICAL ANALYSIS

### A. The tip-sample interaction modeling

Sample surface greatly influences the probe When AFM micro-cantilever becomes close to surface. In the absence of external force, electromagnetic and molecular forces such as Van der Walls forces, capillary forces, and adhesion are main effective interaction forces between sample surface and probe head [17, 18, 19, 20]. Interaction forces play the most important rule in the governing motion of micro cantilever. The interaction forces can be exerted in directions of in-plane and normal directions. Regardless of energy dissipation in tip–sample contact, the normal direction force can be calculated using the JKR [21] or DMT [22] model. The JKR model is applicable for sample with low stiffness material and bigger diameter of tip head while the DMT is suitable for inversely condition. Based on Hertz theorem [23], lateral force ($f_t$) is the function of normal force ($f_n$). The interaction forces between probe and sample can be linearized as visco-elastic model with constant spring coefficient ($k_i$) and constant damping coefficient ($\eta_i$) if the motion of cantilever to be near the equilibrium position with small amplitude [2, 7, 8, 11, 12, 13, 14, 15, 16, 17, 18]. Where normal and tangential forces are:

$$f_n = -k_n(\delta_n - a_n) - \eta_n(\dot{\delta}_n - \dot{a}_n) \tag{1\alpha}$$

$$f_t = -k_t(\delta_t - a_t) - \eta_n(\dot{\delta}_t - \dot{a}_t) \tag{1\beta}$$

Where $f_n$ and $f_t$ represents the normal and tangential linearized interaction forces, $k$ and $\eta$ are the stiffness and damping coefficients of the visco-elastic model, $\delta$ and $a$ are displacement of tip head and displacement of sample surface in normal and tangential directions respectively.

### B. Flexural vibration

The fourth order partial differential equation drives flexural oscillations of the MC with uniform and homogeneous beams and constant cross section. The equation of motion regarding damping coefficients of vertical bending of the MC is [24]:

$$EI_z \frac{\partial^4 y(x,t)}{\partial x^4} + \rho A \frac{\partial^2 y(x,t)}{\partial t^2} + C_v \frac{\partial y(x,t)}{\partial t} = 0 \tag{2}$$

Where $x$ is the coordinate along the longitudinal direction of the micro cantilever, t is time, and $y(x,t)$ is vertical bending of MC, $A$ is cantilever cross section area, $\rho$ is cantilever density, $c_v$ is vertical damping coefficients of MC, $E$ is the Yong's modulus, $I_z$ is the area moment of inertia of MC cross-section.

By assuming harmonic motion of MC holder defined as $h_y(t) = y_0 . e^{i\Omega t}$, the solution of Eq. (2) can be declared by $y(x,t) = Y(x) . e^{i\Omega t}$. Governing equation and corresponding boundary conditions of the MC enable the response function to be obtained which is included in appendix. Due to assuming linear dynamic and harmonic oscillation of MC holder the relative displacement at the tip head ($\delta$) is also assumed in harmonic with the frequency of Ω and amplitude of Δ, where:

$$\delta_i = \Delta_i e^{i\Omega t} \tag{3}$$

The applied forces on the tip head are:

$$f_i = k_i \delta_i + \eta_i \dot{\delta}_i + m_{tip} \ddot{\delta}_t \tag{4}$$

Substituting Eq. 3 into 4 gives:

$$f_i = (k_i + i\eta_i \Omega + m_{tip} \Omega^2) \Delta_i e^{i\Omega t} \tag{5}$$

Regarding above, the consequent applied force is also in harmonic with frequency of Ω. Where,

$$f_i = F_i . e^{i\Omega t} \tag{6}$$

And

$$F_i = (k_i + i\eta_i \Omega + m_{tip} \Omega^2) \Delta_i \tag{7}$$

Regarding above, the governing boundary conditions are:

$$Y(0) = y_0 \tag{8a}$$

$$\frac{dY(x)}{dx}\bigg|_{x=0} = 0 \tag{8b}$$

$$EI \cdot \frac{d^2 Y(x)}{dx^2}\bigg|_{x=L} = F_t \cdot r l_{tip} - F_n \cdot H = (k_t + i\eta_t \Omega + 0.25 m_{tip} \cdot \frac{d^2 Y(x=L)}{dt^2} \Omega^2) \cdot l_{tip} - (k_n + i\eta_n \Omega + m_{tip} \cdot \frac{\partial^2}{\partial t^2} (\frac{dY(x=L)}{dx}) \Omega^2) \cdot H$$

(8c)

$$EI \cdot \frac{d^3 Y(x)}{dx^3}\bigg|_{x=L} = F_t = (k_t + i\eta_t \Omega + m_{tip} \cdot \frac{d^2 Y(x=L)}{dt^2} \Omega^2)$$

(8d)

Eq. 8(a) and 8(b) corresponded to the displacement and slip at $x=0$ respectively. Eq. 8(c) and 8(d) represent momentum and shear force applied at end the of MC ($x=L$).

In Eq. 8(c), $H$ is the sidewall length, $m_{tip}$ is mass of tip, and $l_{tip}$ is the tip length. The conical probe is considered as a concentrated mass where the center of gravity is one-quarter ($r = 0.25$) of the probe length ($l_{tip}$). Thus, the ratio coefficient of $r$ is equal to 1 for applied force of tip-sample interaction and is $1/4$ for the tip mass inertia force. Where $F_n$ and $F_t$ are the incorporation of the visco-elastic interaction forces and the tip mass inertia force in directions of normal and tangential coordinates, respectively.

### III. CASE STUDY AND DISCUSSION

The results of this study were compared with the outcomes and analysis of related researches e.g. [7, 11, 12, 13, 17] and the obtained results are in accordance with mentioned studies. The same properties of MC and its probe were considered as previous studies [13, 17], where the sidewall micro-beam (H) was set 0.5 of MC length. Table 1 illustrates geometric and material parameters of MC and probe. The first and fourth (as a higher RF) vibration modes of MC are studied in this article in which by considering the mentioned properties of MC and its probe the first and fourth RFs was obtained respectively as $\omega_{1,l} = 4.33 \times 10^5 (rad/s)$ and $\omega_{4,l} = 1.03 \times 10^7 (rad/s)$ for Low Interaction Force of tip-sample (LIF), ($k_i = \sim 10^{-6} (N/m)$ and $\eta_i = \sim 10^{-11} (Ns/m)$) and $\omega_{1,h} = 2.47 \times 10^6 (rad/s)$ and $\omega_{4,h} = 1.75 \times 10^7 (rad/s)$ for High Interaction Force of tip-sample (HIF), ($k_i = \sim 10^0 (N/m)$ and $\eta_i = \sim 10^{-5} (Ns/m)$).

TABLE I. PARAMETERS OF SIMULATED AFM CANTILEVER WITH SIDEWALL PROBE

| Cantilever and probe parameters | Magnitude |
|---|---|
| Cantilever length | $234 (\mu m)$ |
| Cantilever Width | $40 (\mu m)$ |
| Cantilever thickness | $3 (\mu m)$ |
| Cantilever density | $2330 (kg/m^3)$ |
| Cantilever Young's modulus ($E$) | $1.5 \times 10^{11} (Pa)$ |
| Cantilever Young's modulus ($G$) | $6.4 \times 10^{10} (Pa)$ |
| Cantilever Vertical damping coefficient ($C_b$) | $8.3 \times 10^{-3} (Ns/m^2)$ |
| Cantilever torsional damping coefficient ($C_t$) | $1.1 \times 10^{-13} (Ns)$ |
| Tip Length | $15 (\mu m)$ |
| Tip mass | $6.54264 \times 10^{-12} (kg)$ |

In order to have no slip condition between the sample and tip and considering the dynamics system of AFM to be linear, the tip-sample distance domain and subsequently the amplitude of the end of MC should not be higher than the critical value during oscillation [13, 17, 25, 26].

Reinstädtler et al. [25, 26] found that a critical domain of excitation amplitude, under a certain applied load, can be determined by detecting the shape of the response resonance curves of the MC. At low excitation amplitudes, the shape of the resonance curve is Lorentzian which with the increasing of excitation amplitude, deviations from the Lorentzian shape appear. Above the critical excitation amplitude, the resonance curve flattens out. The tip-sample interaction affects on the MC amplitude. Hence, the domain of interaction force shall be considered as the meaningful amount. Y. Song et al [13] investigated the interaction force effects, as a visco-elastic force, on the amplitude of end of MC in C-AFMs. The range of interaction force was defined by determining the amount of visco-elastic coefficients of linearized interaction force. Damping and spring coefficients, respectively, were considered about between $\sim 10^{-6}$ to $\sim 10^2$ (N/m) and $\sim 10^{-12}$ to $\sim 10^{-4}$ (Ns/m). In this study, it was found that the meaningful domain of interaction force for AFM-SW is smaller than the C-AFM. This is due to the sidewall micro-beam which causes an increase in applied force and momentum at the end of MC. Indeed, the shear forces and the bending momentum which is applied at the end of MC is greatly larger compared to the C-AFM. It has been discussed that [3, 7, 13, 17] the RF of Common AFM micro-cantilever decreases in LIF condition comparing to HIF. Chang et al. [7] found same result due to AFM-SW yet without considering probe effect. In this paper it is realized that this matter is also true for AFM-SW as it is shown in Fig. 1. However, the tip mass intensifies this decreasing particularly in higher RFs. Hence, it is notable that the increase of RFs due to interaction condition could not be accurate when the tip mass is neglected. For instance, the difference of 4[th] RF between HIF and LIF conditions for considered properties of cantilever is about $0.25 \times 10^7$ rad/s (14% decreasing) without considering tip mass while it is about $0.75 \times 10^7$ rad/s (42.8% decreasing) with considering tip mass as illustrated in Fig. 1 and Fig. 2.

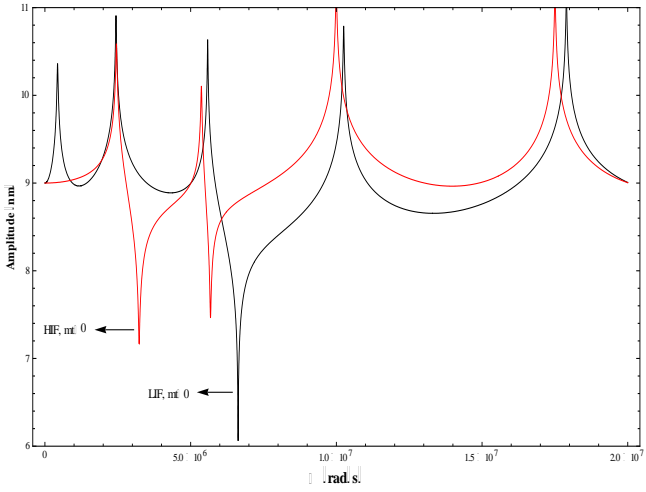

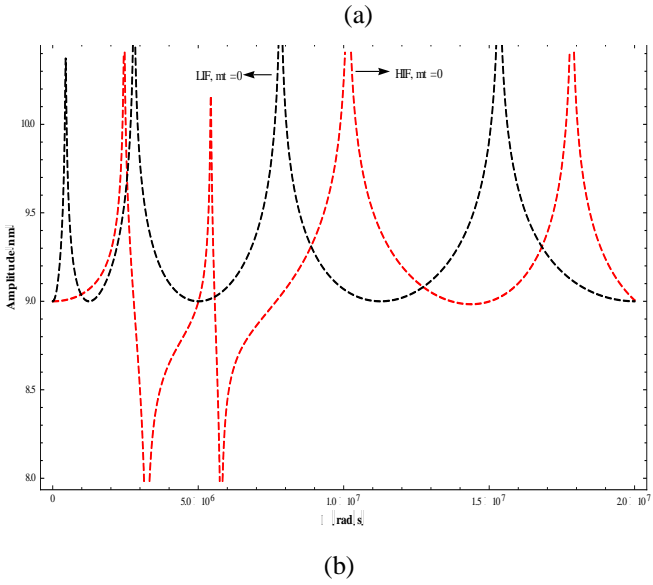

Fig. 1. Micro-cantilever vertical amplitude at Y(L) in cases of (a) Considering and (b) neglecting the tip mass regarding to tip-sample Low Interaction Forces (LIF) and High Interaction Forces (HIF)

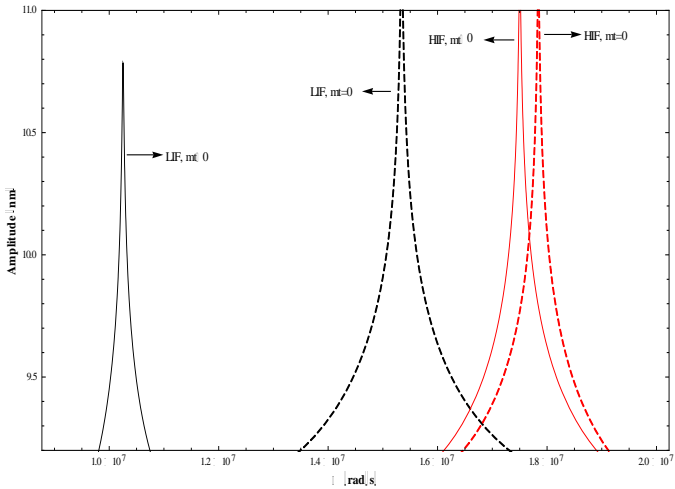

Fig. 2. The 4$^{th}$ resonance frequency of micro-cantilever regarding to tip-sample Interaction Force condition and considering/neglecting the tip mass.

On the other hand, tip mass effect on RFs of Common AFM has been investigated [3, 17]. The results implied that in all situations by considering tip mass, RF decreases compared to ignoring tip mass especially in higher RFs. This means that neglecting the tip mass causes the error in determining RFs whereby the obtained RF is higher than the actual RF. Likewise, this effect happens in AFM-SW. However it is more dominant in AFM-SW compared to C-AFM especially in LIF condition. Results show that the difference of 4th RFs between considering and neglecting the tip mass is about $0.5 \times 10^7$ rad/s (32.7% decreasing) in LIF condition whereas it is about $0.05 \times 10^7$ rad/s (2.8% decreasing) in HIF as shown in Fig. 2 and Fig.3.

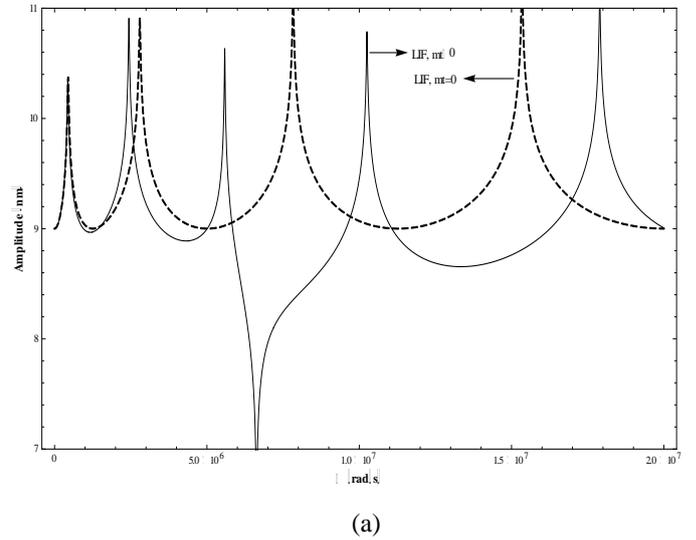

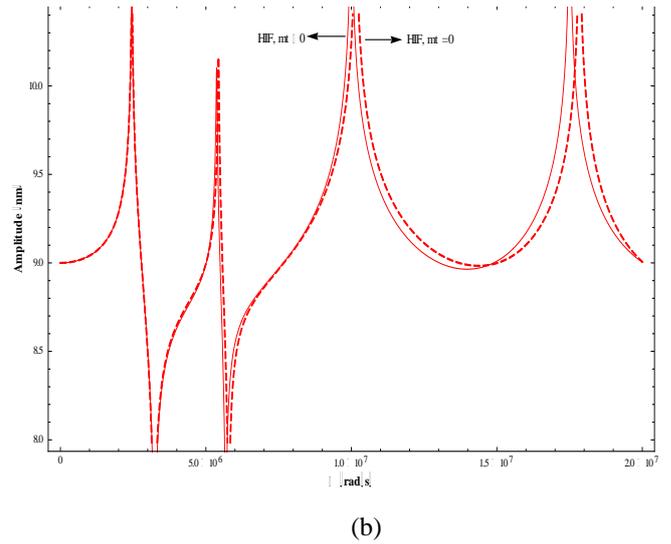

Fig. 3. Micro-cantilever vertical amplitude at Y(L) in cases of (a) tip-sample Low Interaction Forces (LIF) and (b) High Interaction Forces (HIF) regarding to considering/neglecting the tip mass.

It is indicated that the increased in the vertical extension length (H) leads to decrease in the RFs in the case of considering the tip mass, especially in lower interaction force of tip-sample. Fig. 4 illustrates the vertical extension length effect on the MC vertical excitation at $Y(x = L)$ in two cases of LIF and HIF conditions when the sidewall extended micro-beam length is neglected and is equal to half of MC length. Indeed, probe mass effect on RF, which causes dropping of RF, will be magnified when sidewall beam is longer.

(H) is longer, but by decreasing interaction force, sidewall micro-beam length effect on amplitude remains almost unaffected.

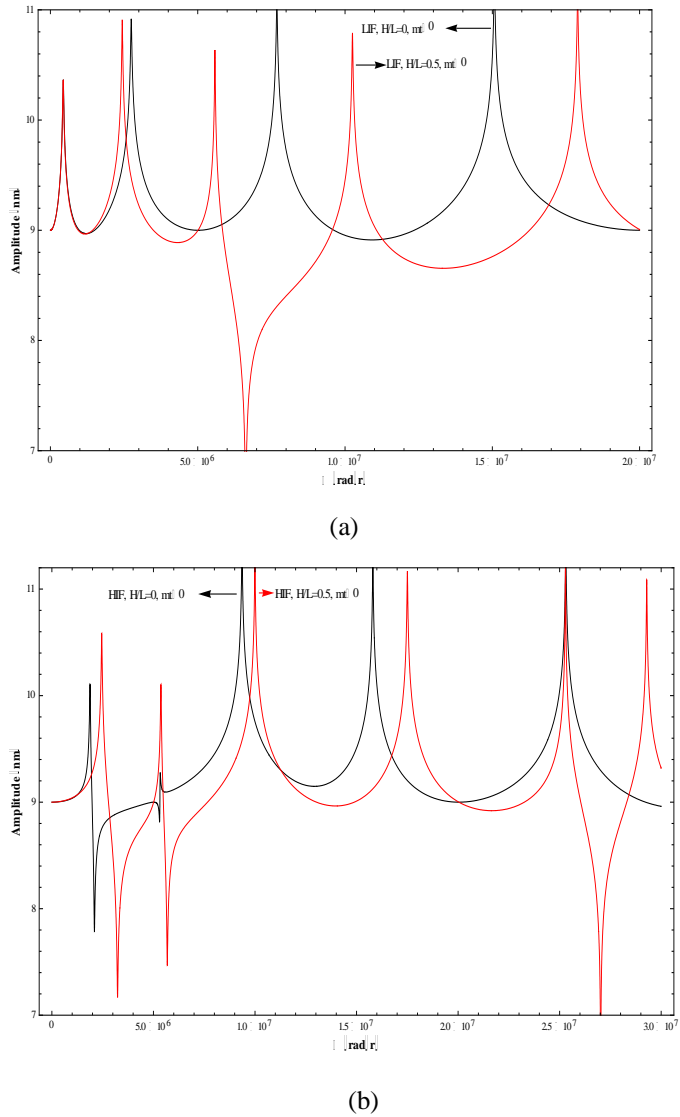

(a)

(b)

Fig. 4. Sidewall beam length influence on Resonance Frequencies in cases of (a) LIF and (b) HIF situations regarding to considering/neglecting the sidewall beam length

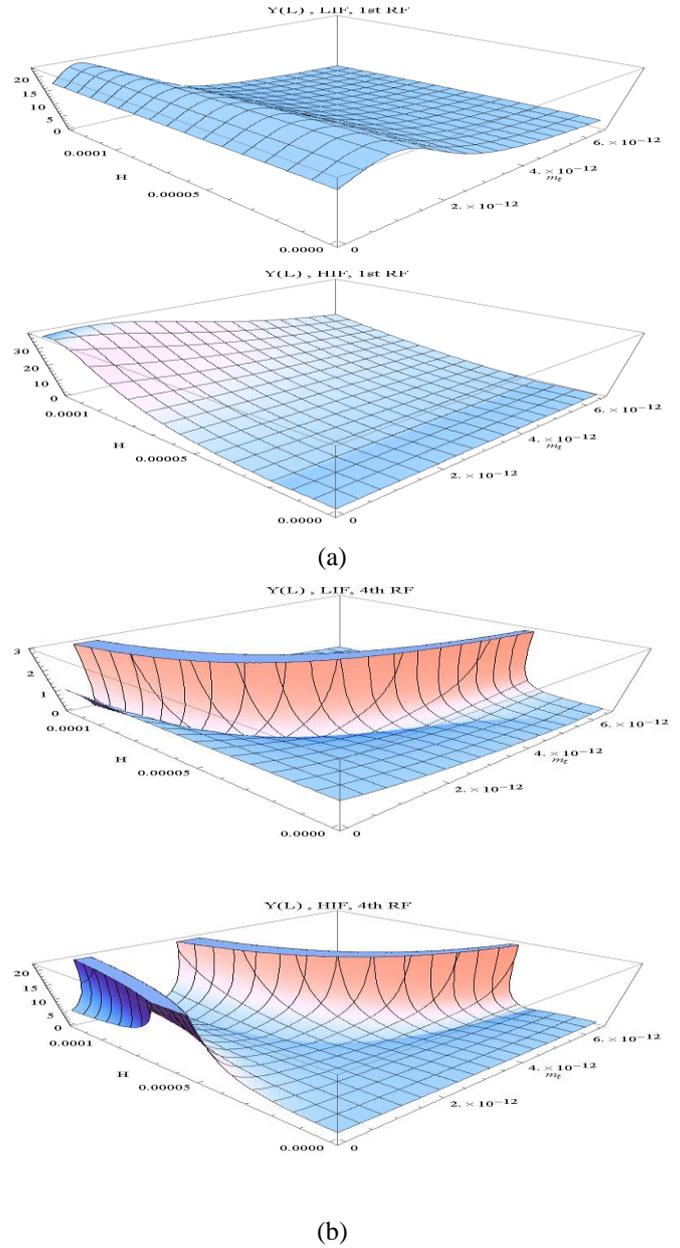

(a)

(b)

Fig. 5. Micro-cantilever amplitude responses at Y(L) to unit excitation amplitude $Y_0$ (a) in First RF and (b) in higher RF in cases of LIF and HIF, regarding to the tip mass and the height of sidewall micro-beam

Generally, the MC deflection and amplitude are utilized to analyze the sample surface [2]. However, reciprocal effects of various sets of parameters on MC, due to stiffness of MC, inertia force, bending moment of inertia of probe mass, and tip-sample interaction, cause complicated recognizing of the MC deflections. Hence, the study of MC deflections is required. 3D Fig. 5 illustrates vertical amplitude at the end of MC ($Y(x=L)$) regarding the tip mass ($0-6.54264\times10^{-6}$ kg) and the sidewall micro-beam length ($0-117\times10^{-6}$) in two different conditions of LIF ($k_i \cong 10^{-6}, \eta_i \cong 10^{-11}$) and HIF ($k_i \cong 10^0, \eta_i \cong 10^{-6}$) in 1st and 4th RFs ($\omega_{1,l}$, $\omega_{1,h}$, $\omega_{4,l}$ and $\omega_{4,h}$). In first RF, it is shown that the vertical amplitude decreases, when tip mass is heavier. Besides, amplitude increases if sidewall micro-beam

The MC amplitude of first and fourth modes vibration which is obtained by the probe mass ($0-6.54264\times10^{-12} m$) and the sidewall extension length ($0-234\times10^{-6} m$) in case of considering probe mass are shown in Fig. 6 and 7 respectively. The reciprocal effects of inertia force and bending moment of inertia of probe mass lead to tolerance in MC deflections. Considering figures, the noticeable tolerance as a peak in amplitude of MC is prominent especially in higher RF which is clear for 4th RF in Figures. The results obtained from considered properties of MC and its probe show us that the pick of tolerance occurs in the half length of MC ($234 \times 10^{-6} m$). Mokhtari-Nezhad et al. [17] also detected the peak tolerance of MC amplitudes that can happen in the specific range of tip mass. Fig. 7 shows these peak tolerances of vertical amplitude at $Y(x = L)$ in first and higher RFs. However, it is generally noticeable that in general, by the increase of tip mass, the vertical MC amplitude finally diminishes in all situations.

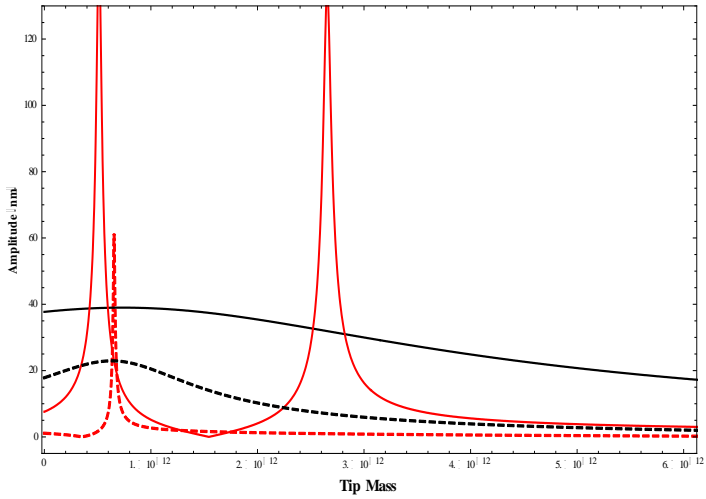

Fig. 6. The mico-cantilever amplitude responses at Y(L) to unit excitation amplitude $Y_0$ in 1st and 4th RF regarding to the HIF and LIF condition

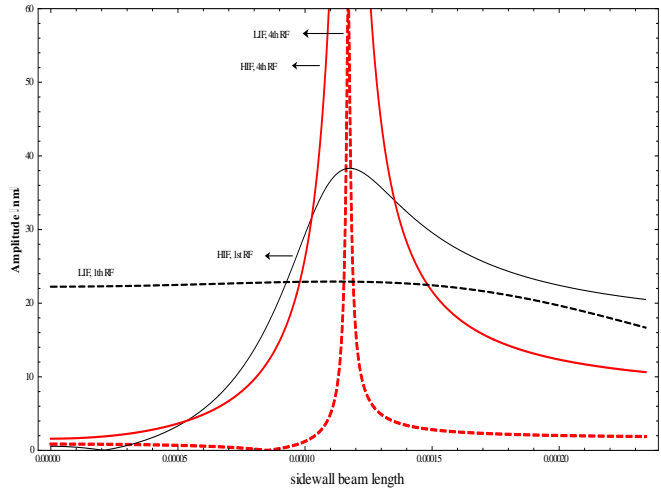

Fig. 7. The mico-cantilever amplitude responses at Y(L) to unit excitation amplitude $Y_0$ in 1st and 4th RF by considering probe mass regarding to the sidewall extension length

## IV. CONCLUSION

In this paper, an analytical study has been developed for expressing the dynamics behavior of MC and its probe system in AFM with SideWall probe (AFM-SW). In addition, the effects of interaction force, tip mass, and sidewall micro-beam length, on vibration and deflection of MC have been widely studied. It has been demonstrated that the variations of the interaction force and the tip mass cause the shifting of the RF. Besides, the RF shifting in AFM-SW micro-cantilever is observed to be more considerable in comparison with the Common AFM. The tip mass causes reduction of the RFs especially in higher RFs and weaker tip-sample interaction. Moreover, the tip mass effects are more dominant for higher length of sidewall micro-beam. In addition, probe mass and sidewall length effects on MC amplitude has been studied. Noticeable tolerances as a pick occur in MC amplitude in specific range of probe mass and sidewall length specially in higher RFs. It is shown that, generally, the heavier tip mass causes to diminish the amplitude of the end of MC. Although the analytical results are well match with the previous studies, these results outperform the previous ones in the AFM-SW dynamics point of view.

## V. APPENDIX

Response function of MC of AFM-SW to excitation amplitude $Y_0$ in Vertical Excitation mode in order to considering probe mass:

$Y(L)=Y_0(H^2(k_n+\omega(ic_n-m_{tip}\omega))((1+e^{2ia_vL}-4e^{(1+i)a_vL}+e^{2a_vL}+e^{(2+2i)a_vL})k_l+2a_v^3(e^{ia_vL}-ie^{a_vL}+ie^{(1+2i)a_vL}-e^{(2+i)a_vL})EI_z+i(1+e^{2ia_vL}-4e^{(1+i)a_vL}+e^{2a_vL}+e^{(2+2i)a_vL})\omega(c_l+im_{tip}\omega))-a_vEI_z((ia_vl_{tip}(-1+e^{2ia_vL})(-1+e^{2a_vL})-(1+i)(-1+ie^{2ia_vL}-ie^{2a_vL}+e^{(2+2i)a_vL}))k_l+2a_v^3(e^{ia_vL}+e^{a_vL}+e^{(1+2i)a_vL}+e^{(2+i)a_vL})EI_z-\frac{1}{3}ia_vl_{tip}(-1+e^{2ia_vL})(-1+e^{2a_vL})m_{tip}\omega^2+(1+i)(-1+ie^{2ia_vL}-ie^{2a_vL}+e^{(2+2i)a_vL})\omega(-ic_l+m_{tip}\omega))))/(H^2(k_n+\omega(ic_n-m_{tip}\omega))((1+e^{2ia_vL}-4e^{(1+i)a_vL}+e^{2a_vL}+e^{(2+2i)a_vL})k_l+(1+i)a_v^3(-i+e^{2ia_vL}-e^{2a_vL}+ie^{(2+2i)a_vL})EI_z+i(1+e^{2ia_vL}-4e^{(1+i)a_vL}+e^{2a_vL}+e^{(2+2i)a_vL})\omega(c_l+im_{tip}\omega))-a_vEI_z((ia_vl_{tip}(-1+e^{2ia_vL})(-1+e^{2a_vL})-(1+i)(-1+ie^{2ia_vL}-ie^{2a_vL}+e^{(2+2i)a_vL}))k_l+a_v^3(1+e^{2ia_vL}+4e^{(1+i)a_vL}+e^{2a_vL}+e^{(2+2i)a_vL})EI_z-a_vl_{tip}(-1+e^{2ia_vL})(-1+e^{2a_vL})\omega(c_l+\frac{im_{tip}\omega}{3})+(1+i)(-1+ie^{2ia_vL}-ie^{2a_vL}+e^{(2+2i)a_vL})\omega(-ic_l+m_{tip}\omega))$

Where,

$a_v=[\frac{\rho A}{EI_z}\omega^2-i\frac{c_v}{EI_z}\omega]^{1/4}$